\documentclass[reprint,prl,twocolumn,twoside,aps,superscriptaddress]{revtex4-2}

\usepackage{amsmath,amssymb,amsfonts}
\usepackage{graphicx}
\usepackage{siunitx}
\usepackage{natbib}
\usepackage{braket}
\usepackage{multirow}
\usepackage[normalem]{ulem}
\usepackage[dvipsnames]{xcolor}
\usepackage{xspace}
\usepackage{url}
\usepackage[breaklinks]{hyperref}
\usepackage{braket}

\usepackage[on]{svg-extract} % view saved files by going to other logs and files in output
\usepackage{svg}
\svgsetup{inkscapearea=page,pretex=\small,clean=true}
\svgpath{{svgs/}}

\hypersetup{
    unicode=true,
    colorlinks=true,
    linkcolor=blue,
    citecolor=blue,
    urlcolor=blue
  }
\usepackage{breakurl}

\newcommand{\one}{$\mathrm{S}$\xspace}%{$\left(-1 (1, 3) \textrm{s} (1, 3)\right)$\xspace}
\newcommand{\onespinflip}{S$'$\xspace}%{$\left(-1 (2, 3) \textrm{s} (2, 3)\right)$\xspace}
\newcommand{\g}{$\mathrm{G}$\xspace}%{$\left(-1 (1, 3) \textrm{g} (1, 3)\right)$\xspace}
\newcommand{\two}{$\mathrm{D}$\xspace}%{$\left(-2 (1, 3) \textrm{d} (0, 3)\right)$\xspace}
\newcommand{\intermediate}{$(^{3}\Pi_{1}, v'=29, J'=1)$\xspace}

\usepackage{xr}
\begin{document}

\title{Formation of ultracold molecules by merging optical tweezers}

% author issues fixed by specifiying superscriptaddress in documentclass
\newcommand{\physics}{Department of Physics and Joint Quantum Centre (JQC) Durham-Newcastle, Durham University, South Road, Durham, DH1 3LE, United Kingdom}
\newcommand{\chemistry}{Department of Chemistry and Joint Quantum Centre (JQC) Durham-Newcastle, Durham University, South Road, Durham, DH1 3LE, United Kingdom}

\author{Daniel K. Ruttley}
\thanks{D. K. R. and A. G. contributed equally to this work.}
\affiliation{\physics}
\author{Alexander Guttridge}
\thanks{D. K. R. and A. G. contributed equally to this work.}
\affiliation{\physics}
\author{Stefan Spence}
\affiliation{\physics}
\author{Robert~C.~Bird}
\affiliation{\chemistry}
\author{C. Ruth Le Sueur}
\affiliation{\chemistry}
\author{Jeremy M. Hutson}
\email{j.m.hutson@durham.ac.uk}
\affiliation{\chemistry}
\author{Simon L. Cornish}
\email{s.l.cornish@durham.ac.uk}
\affiliation{\physics}

\begin{abstract}
We demonstrate the formation of a single RbCs molecule during the merging of two optical tweezers, one containing a single Rb atom and the other a single Cs atom. Both atoms are initially predominantly in  the motional ground states of their respective tweezers. We confirm molecule formation and establish the state of the molecule formed by measuring its binding energy. We find that the probability of molecule formation can be controlled by tuning the confinement of the traps during the merging process, in good agreement with coupled-channel calculations. We show that the conversion efficiency from atoms to molecules using this technique is comparable to magnetoassociation. 

\end{abstract}
\date{\today}

\maketitle
Arrays of molecules confined in optical potentials are a powerful platform for quantum science \cite{Kaufman2021}. Full quantum state control of molecules is essential to realise the potential of this system in the domains of quantum simulation \cite{Barnett2006,Gorshkov2011,Baranov2012,Lechner2013,Wall2015,Bohn2017,Yao2018} and quantum information processing \cite{DeMille2002,Ni2018,Hughes2020,Albert2020,Sawant2020}. 
Recent progress on trapping molecules in optical tweezers and optical lattices has demonstrated the first steps toward this goal, including single-site readout of individual molecules \cite{Anderegg2019}, entanglement of pairs of molecules \cite{Christakis2023,Holland2022,Bao2022} and the assembly of molecules predominantly occupying a single motional level of an optical tweezer \cite{Cairncross2021}.

Ultracold molecules may be prepared in optical potentials by following an indirect approach, where the molecules are produced by associating pre-cooled pairs of atoms. This approach benefits from the wealth of techniques developed for the cooling and control of atoms and results in the formation of molecules which inherit the low temperatures of the constituent atoms. 
The formation of molecules often follows a two-step process where first weakly bound molecules are produced from an atomic sample using magnetoassociation \cite{Koehler2006,Chin2010} and then the weakly bound molecules are transferred to the rovibrational ground state using stimulated Raman adiabatic passage (STIRAP) \cite{Bergmann1998,Vitanov2017}. Magnetoassociation  exploits an avoided crossing between atomic and molecular states as a function of magnetic field and has been widely employed to convert atoms trapped in weakly confining optical potentials into molecules. 

When the atoms are trapped in the tightly confining potentials of optical tweezers or lattices, only atom pairs in the relative motional ground state of the optical trap can be converted into weakly bound molecules using magnetoassociation \cite{Busch1998}. In addition, confinement-related effects arise when the harmonic confinement length approaches the value of the s-wave scattering length. Effects like elastic \cite{Olshanii1998,Bergeman2003} and inelastic \cite{Kestner2010,Sala2016} confinement-induced resonances (CIRs) have been observed experimentally in a number of different systems and dimensionalities \cite{Moritz2005,Haller2009,Lamporesi2010a,Zuern2012,Capecchi2022}. These confinement-related effects offer new ways to form molecules. Using pairs of fermions in 1D, inelastic CIRs have been used to form molecules coherently in an optical trap \cite{Sala2013}. In addition, molecules have been formed coherently utilising spin-motion coupling in a strongly focused optical tweezer with large polarisation gradients \cite{He2020}.

\begin{figure}[t] 
\includegraphics[width=\columnwidth]{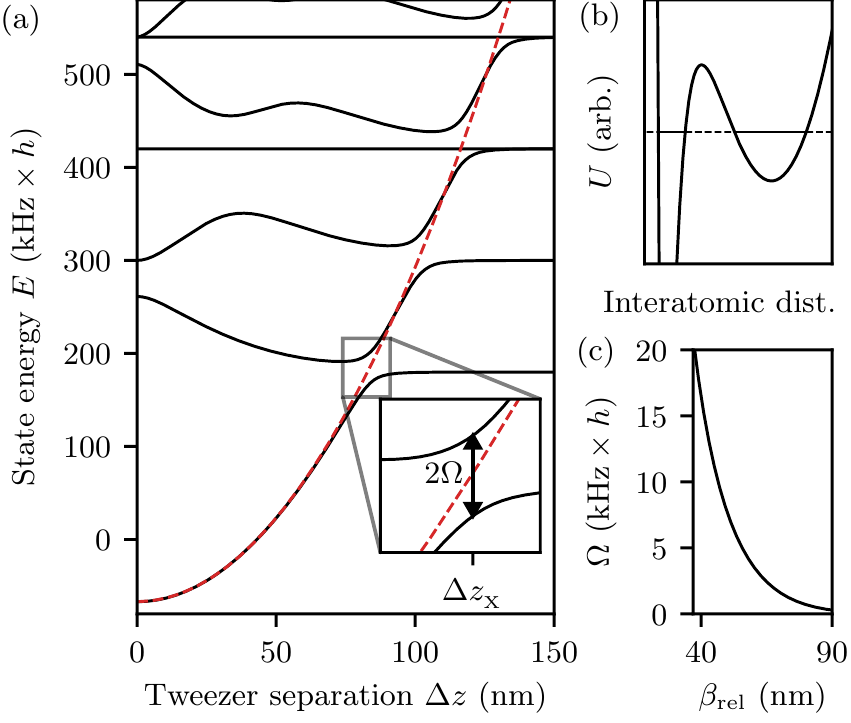}
\caption{(a) Energy levels for the Rb+Cs system as a function of separation between two spherically symmetric optical tweezers with confinement length $\beta_\textrm{rel} = 40$~nm for relative motion. The red dashed line shows the harmonic trapping experienced by the molecule in the absence of tunnelling. An avoided crossing between the molecular and atom-pair states at $\Delta z_{_{\textrm{X}}}$, expanded in the inset, allows mergoassociation. (b) Cartoon of the interaction energy as a function of interatomic distance when $\Delta z = \Delta z_{_{\textrm{X}}}$. 
(c) The effective matrix element $\Omega$ between the lowest-energy atom-pair and molecular states as a function of $\beta_\textrm{rel}$.}
\label{fig:theory}
\end{figure}

 In contrast, there has been little experimental investigation of the interactions of two particles in separate 
optical potentials with tuneable separation \cite{Kaufman2014,Kaufman2015}, despite the existence of theoretical work in this area \cite{Stock2003,Krych2009}.  
Stock \textit{et al.} \cite{Stock2003} predicted the existence of avoided crossings between molecular and confined-atom states at critical values of the separation of two optical potentials. They termed these trap-induced shape resonances.
Figure~\ref{fig:theory}(a) shows the energies of a system with two atoms in separate but overlapping traps as a function of trap separation $\Delta z$. At large separation, the energies of the separately confined atom pairs are almost independent of $\Delta z$. However, there can also be a molecular state that is weakly bound at $\Delta z=0$. The energy of this state increases quadratically with $\Delta z$ due to the tweezer potentials and there is an avoided crossing with the lowest confined-atom state at a critical separation $\Delta z_{_{\textrm{X}}}$.  The strength of the avoided crossing depends on the height and width of the barrier between the atomic and molecular wells, as shown in Fig.~\ref{fig:theory}(b); it is greatest when 
the bound state is close to threshold, corresponding to a large positive value of the s-wave scattering length $a_\textrm{s}$, and when the confinement length $\beta_\textrm{rel}$  for relative motion of the atoms is comparable to $a_\textrm{s}$. This avoided crossing offers an unexplored path to the formation of molecules by merging together two optical potentials. 
This was not observed in previous demonstrations of molecule formation in lattices \cite{Covey2016,Reichsoellner2017,Christakis2023} and tweezers \cite{Zhang2020,He2020}, probably because $a_\textrm{s}\ll\beta_\textrm{rel}$ for the systems investigated.

In this Letter, we report the observation of molecule formation through merging of two optical tweezers, one containing a single Rb atom and the other a single Cs atom.  Guided by coupled-channel calculations, we elucidate the experimental conditions that are necessary for molecule formation by merging of the optical potentials, a process referred to here as \emph{mergo}association. 
We demonstrate mergoassociation by using optical spectroscopy to measure the binding energy of the molecular state occupied after molecule formation. The interaction potential between Rb and Cs atoms is accurately known \cite{Takekoshi2012} and comparison of our measurements with coupled-channel calculations of the near-threshold bound states allows us to identify the molecular state and  understand the molecule formation process.  We explore the tunability of the molecule formation probability by controlling the confinement strength during merging and compare the molecule formation probability to that obtained using magnetoassociation.
Finally, we confirm that mergoassociation can be performed at low magnetic fields, without any magnetic field ramps, and detect the formation of molecules using microwave spectroscopy. This demonstrates the utility of mergoassociation in systems that do not possess Feshbach resonances suitable for magnetoassociation.

Our experiments begin by preparing single \(^{87}\)Rb and \(^{133}\)Cs atoms in hyperfine states $\left(f_\textrm{Rb} = 1, m_\textrm{Rb} = 1\right)$ and $\left(f_\textrm{Cs} = 3, m_\textrm{Cs} = 3\right)$ in the motional ground states of spatially separated, species-specific optical tweezers \cite{Spence22,supplement}. 
These tweezers are subsequently brought together in order to form a molecule.
The Rb+Cs atom pair state $(1,1)+(3,3)$ has a near-threshold bound state with binding energy $110\pm2~\mathrm{ kHz}\times h$ at magnetic fields far from any Feshbach resonance; this corresponds to an interspecies background scattering length $a_\textrm{s} = 645(60)\ a_0$ (approx. 34.1(3)~nm) \cite{Takekoshi2012}. 
The binding energy is comparable to the energy spacing of the harmonic levels in the tweezers and therefore within the regime where Stock \textit{et al.} \cite{Stock2003} predicted the existence of strong avoided crossings.

We have carried out coupled-channel calculations of the energy levels for pairs of atoms in separated tweezers, using the methods described in Supplemental Material \cite{supplement}. Our calculations treat the individual tweezers as spherical and harmonic and neglect coupling between motions in the relative and center-of-mass coordinates. 
We represent the atom-atom interaction with a point-contact potential chosen to reproduce $a_\textrm{s}$. The resulting energies for $\beta_\textrm{rel}=40$~nm are shown in Fig.~\ref{fig:theory}(a). For each value of $\beta_\textrm{rel}$, we locate $\Delta z_{_{\textrm{X}}}$ and characterize the strength of the avoided crossing in terms of an effective matrix element $\Omega$; this gives the strengths shown in Fig.~\ref{fig:theory}(c). The value of $\Delta z_{_{\textrm{X}}}$ is approximately $\beta_\textrm{rel} \sqrt{3 + \beta_\textrm{rel}^{2} / a_\textrm{s}^{2}}$, so the crossing occurs at larger interatomic separations for weaker confinement. This leads to a reduction in tunneling through the barrier and in the strength of the avoided crossing.

If the avoided crossing is sufficiently strong, it may be traversed adiabatically with a slow enough change in $\Delta z$. This leads to the conversion of an atom pair in the ground state of relative motion into a molecular bound state. We calculate the probability of traversing the avoided crossing adiabatically using  Landau-Zener theory 
 \cite{supplement}.

\begin{figure}[t]
\includegraphics[width=\columnwidth]{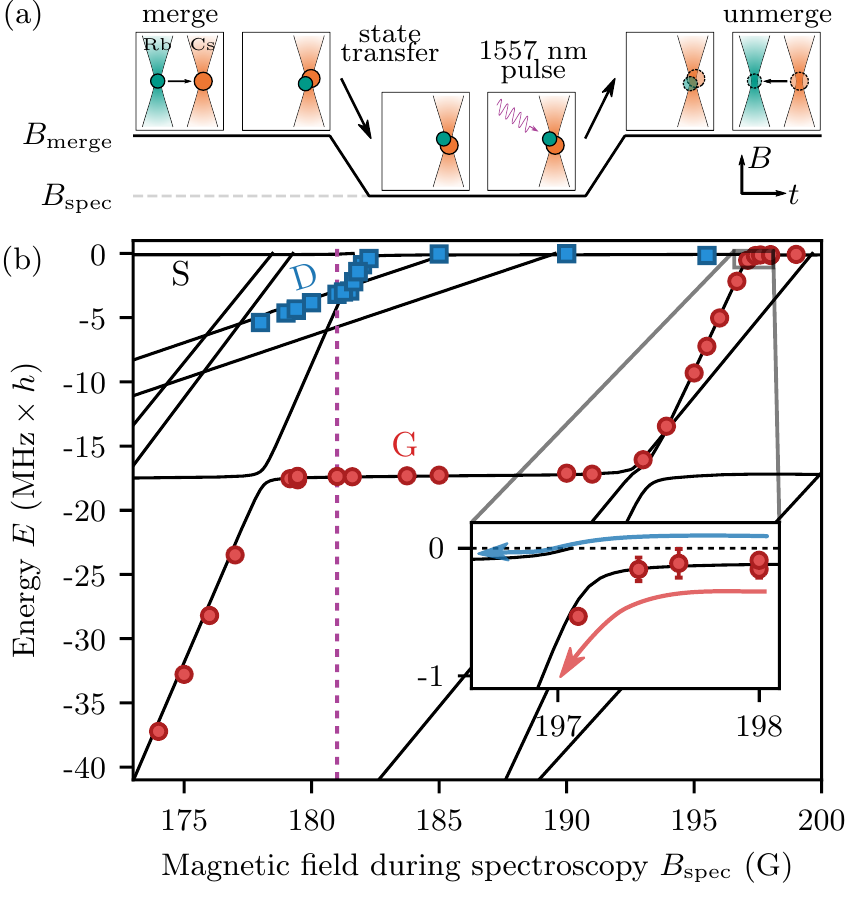}
\caption{(a) Sequence for molecule formation and detection. (b) The energy of weakly bound RbCs molecules (relative to threshold) as a function of magnetic field, $B_{\rm{spec}}$. The points show the measured energies of molecules produced when merging the traps at 205~G (red circles) and below 197~G (blue squares). The purple dashed line indicates the field used for spectroscopy in Fig.~\ref{fig:mergo} from states D and G. Black lines show state energies calculated from the RbCs molecular potential of Ref.~\cite{Takekoshi2012}. The inset shows the avoided crossing below the resonance near 197~G and highlights the paths for mergoassociation (red arrow) and magnetoassociation (blue arrow)~\cite{Koeppinger2014}.
\label{fig:1557}}
\end{figure}

Figure~\ref{fig:1557}(a) shows the experimental sequence used to probe and dissociate molecules.
A single Rb atom and a single Cs atom are prepared in species-specific tweezers which are merged together at magnetic field $B_\textrm{merge}$. The field is then ramped down to $B_\textrm{spec}$.
Atom pairs can either be mergoassociated during the merging step if the tweezer confinement is sufficiently strong or magnetoassociated if the magnetic field ramp crosses a Feshbach resonance.
A spectroscopy pulse of light at $1557$~nm is applied before the field ramps are reversed and the traps are unmerged.
Then, fluorescence imaging of the atoms is performed to determine the trap occupancy.
When the spectroscopy light is resonant with a molecular transition, loss to other molecular states results in no atoms being reimaged \cite{Debatin2011,Molony2014,Takekoshi2014,Molony2016a}. 
We identify the molecular states that have been populated during a sequence by comparing to coupled-channel calculations using the RbCs interaction potential of Takekoshi \emph{et al.} \cite{Takekoshi2012,supplement}.

Figure~\ref{fig:1557}(b) shows the binding energies of RbCs molecules formed by mergoassociation, measured using the optical spectroscopy \cite{supplement}.
Molecules formed when merging the traps at $B_\mathrm{merge} = 205$~G follow the path indicated by red circles when the field is ramped to $B_\mathrm{spec}$.
Entry into the near-threshold bound state by mergoassociation above the Feshbach resonance at $197.08(2)$~G allows us to approach this resonance from the molecular side, as shown in the inset,  and subsequently to occupy states not accessible in magnetoassociation experiments from this starting field \cite{Takekoshi2014,Koeppinger2014}.
By mergoassociating with $B_\mathrm{merge}$ below this resonance we instead follow the path indicated by blue square points  as the field is ramped.
For the purposes of this investigation, we utilise the difference in energy between the states \g and \two at $B_\mathrm{spec} = 181$~G (purple dashed line) to distinguish between molecules that have followed these two paths.

\begin{figure}[t] 
\includegraphics[width=\columnwidth]{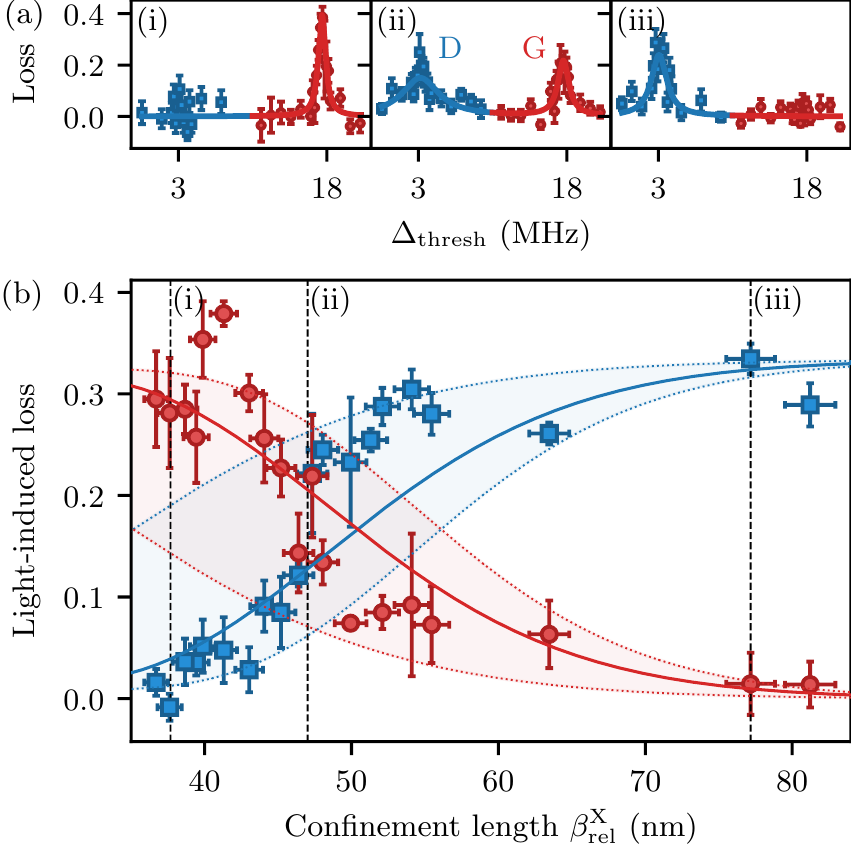}
\caption{(a) Spectroscopic identification of molecules formed by mergoassociation (red circles) and magnetoassociation (blue squares) for confinement lengths $\beta_\textrm{rel}^\textrm{X}$ at the avoided crossing of (i) 37.7(8)~nm, (ii) 47(1)~nm, and (iii) 77(2)~nm. Light-induced loss is measured as a function of the detuning $\Delta_\mathrm{thresh}$ from the transition between the atomic threshold and \intermediate at 181~G. 
(b) The probability of light-induced loss of molecules formed by mergoassociation (red circles) and magnetoassociation (blue squares) as a function of $\beta_\textrm{rel}^\textrm{X}$.
The theory curves show the calculated Landau-Zener probabilities scaled to match the light-induced loss. 
The shaded regions indicate the experimental uncertainty in the merging speed.
\label{fig:mergo}}
\end{figure} 

Figure~\ref{fig:mergo} shows how the probability of mergoassociation depends on the tweezer confinement length $\beta_\textrm{rel}^\textrm{X}$ at the avoided crossing.
For these measurements, the traps are merged at an average speed $1.4$~\textmu m/ms. However, the speed $(d\Delta z/dt)_\mathrm{X}$ at the avoided crossing depends strongly on the alignment of the two tweezers; simulations of the combined potential predict that $(d\Delta z/dt)_\mathrm{X} = 0.9^{+2.7}_{-0.4}$~\textmu m/ms \cite{supplement}.
During the merging step, the applied magnetic field is $B_\mathrm{merge} = 205$~G.
Following merging, the magnetic field is jumped down to $199$~G and then ramped down to $196.8$~G in 3~ms; this step magnetoassociates any remaining atom pairs in the relative motional ground state with an expected conversion efficiency greater than 99\%. 
The magnetic field is then ramped down to $B_\mathrm{spec} = 181$~G in 3~ms in order to perform spectroscopy. 
The frequency of the spectroscopy laser at which we observe light-induced loss allows us to determine whether the molecule occupies state \g or \two and hence whether it was formed by mergoassociation or magnetoassociation. The panels in Fig.~\ref{fig:mergo}(a)(i)-(iii) show optical spectra for strong, intermediate, and weak confinement during merging.
For strong confinement we observe high  occupation of state \g as the majority of atom pairs in the motional ground state are mergoassociated.
As the confinement is reduced, fewer atom pairs are mergoassociated, resulting in high occupation of state~\two.

 Figure~\ref{fig:mergo}(b) shows the probability of excitation by the spectroscopy light as a function of $\beta_\textrm{rel}^\textrm{X}$ during merging of the traps.
The spectroscopy laser saturates the transition from either \g or \two to determine the occupied molecular state.
We clearly observe a change in the probability of mergoassociation from high to low as the confinement is reduced. 
The peak probability of the mergoassociation data points (red circles) and the magnetoassociation data points (blue squares) indicates that the efficiency of  the two techniques is similar. 
The red (blue) curve shows the calculated Landau-Zener probability $P_\textrm{LZ}$ \cite{supplement}  of mergoassociation (no mergoassociation) for our experimental parameters, scaled to that of the light-induced loss.
The shaded regions show the uncertainty in $P_\textrm{LZ}$ arising from the uncertainty in $(d\Delta z/dt)_\mathrm{X}$. 
The experimental results are in good agreement with our theoretical model, with the observed crossover point in $\beta_\textrm{rel}$ within $10\%$ of the theoretical prediction. 
The agreement is surprisingly good in view of the approximations made in the model, particularly the assumption that the tweezers are spherically symmetric. 
In reality, our tweezers have an aspect ratio of 2:5 between the confinement lengths along the axes of approach and tweezer-light propagation. 
 Both magnetoassociation and mergoassociation rely on the initial atom pair residing in the ground state of relative motion, and this state preparation is the current limit to our association efficiency.

\begin{figure}[t]
\includegraphics[width=\columnwidth]{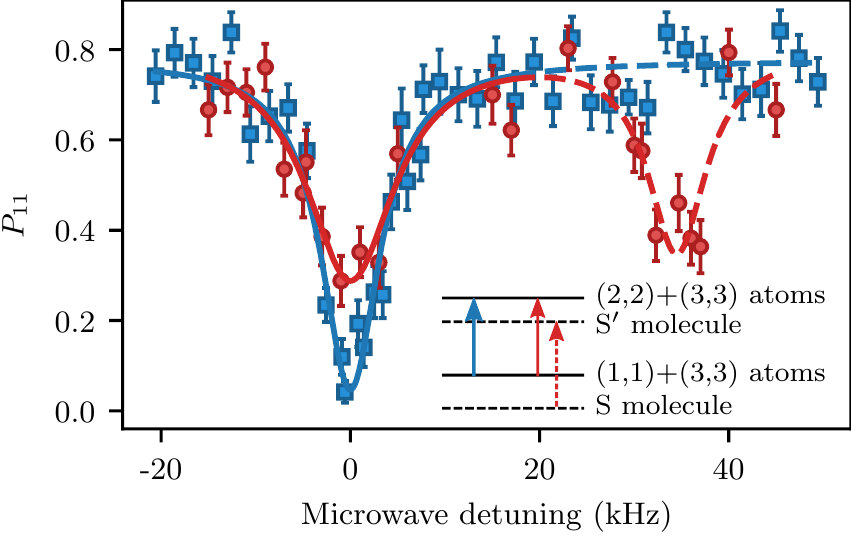}
\caption{Microwave spectroscopy of RbCs molecules produced by mergoassociation at $B=4.78$~G.  The probability $P_{11}$ of detecting both a Rb and Cs atom at the end of the sequence is shown as a function of the microwave detuning from the transition $(1,1)\rightarrow(2,2)$ in atomic Rb for strong (red circles) and weak (blue squares) confinement during merging: $\beta_\mathrm{rel}^\mathrm{X} = 39.4(9)$~nm and $55(1)$~nm respectively. When a molecule is formed, we observe the molecular transition {\one$\rightarrow$\xspace\onespinflip} at detuning $35(2)$~kHz. The inset shows the energy-level structure of the relevant states with atomic (molecular) states indicated with solid (dashed) lines. 
\label{fig:MW}}
\end{figure}

Finally, we verify that mergoassociation can be performed at low magnetic fields, without any magnetic field ramps.
Mergoassociation is still possible in this regime, because it relies only on the presence of a bound state near threshold, which for RbCs exists over a large range of magnetic field. 
The experimental sequence is similar to the one described earlier, but the magnetic field is held constant at the field applied during cooling of the atoms, $B= 4.78$~G, for the entirety of the molecule formation and detection portion of the sequence.
A microwave pulse of frequency $\sim 6.84$~GHz is applied for $89$~\textmu s in place of the pulse of light at 1557~nm. This pulse length approximates a $\pi$-pulse for the Rb atom and for the RbCs molecule in the least-bound state \one \cite{supplement}.
Following the unmerging of the traps, a resonant ``pushout" pulse is applied;  this ejects any Rb atoms in the state {$\left(f=2\right)$ \cite{Brooks21} to allow state-sensitive detection of the Rb atom.

The results of the microwave spectroscopy are  shown in Fig.~\ref{fig:MW}. 
The blue squares show the results for weak confinement during merging, where the probability of mergoassociation is low and we expect to prepare an atom pair.
We observe only a single feature in the probability $P_{11}$ of observing a Rb and Cs atom at the end of the sequence; this is the feature corresponding to the hyperfine transition $(1,1)\rightarrow(2,2)$ in atomic Rb.
In contrast, when the tweezers are merged with stronger confinement, as shown by the red circles, we mergoassociate the atom pair to form a molecule. 
Consequently, we observe an additional feature in $P_{11}$, detuned by $35(2)$~kHz; this corresponds to the molecular transition {\one$\rightarrow$\xspace\onespinflip} illustrated in the inset.
Both features are fitted with Lorentzians as shown by solid (dashed) lines for the atomic (molecular) transition. Using the RbCs interaction potential fitted in Ref.~\cite{Takekoshi2012}, we calculate the binding energy of state \onespinflip at 4.78~G to be $80~\mathrm{ kHz}\times h$. This value is smaller than that of \one ($122~\mathrm{ kHz}\times h$) and the calculated difference in binding energy ($42~\mathrm{ kHz}\times h$) is in reasonably good agreement with the experimental measurement. 
The atom pair is prepared in the required hyperfine states in 78(1)\% of experimental runs and the relative depths of the features in Fig.~\ref{fig:MW} indicate that $46(8)$\% of these atom pairs are converted into molecules.

In summary, we have created a trap-induced avoided crossing between atomic and molecular states and used it to create molecules during the merging of pairs of optical tweezers. The efficiency of molecule formation depends on the strength of the avoided crossing, which critically depends on the  confinement length for relative motion. 
The avoided crossing is strongest when there is a bound state near threshold and the confinement length is comparable to the scattering length. This situation is realised for Rb+Cs at a large range of magnetic fields. We have demonstrated that the efficiency of molecule formation by mergoassociation is comparable to that of magnetoassociation in this system.

This work demonstrates a new technique for the formation of molecules in systems with large interspecies interactions. It will be effective even in systems that do not possess Feshbach resonances suitable for magnetoassociation \cite{Guttridge2018a}. 
It would be interesting to test this technique using transport in an optical lattice, where the tighter confinement achievable should allow efficient conversion of atom pairs into molecules for systems with moderate positive scattering lengths. This would also open up applications in neutral-atom quantum computing by using the trap-induced avoided crossing for high-fidelity two-qubit quantum logic operations \cite{Stock2006}. Further, the observation of such features in the merging of tweezers has important ramifications for collision measurements using tweezer-confined particles.

This work also demonstrates the first production of RbCs molecules in optical tweezers. These weakly bound molecules can be transferred to the rovibrational ground state using STIRAP as has been previously demonstrated for weakly confined samples of RbCs \cite{Molony2014,Takekoshi2014,Molony2016a}.
The production of ground-state molecules using these techniques prepares the molecules predominantly in the lowest motional state of the trap, which will allow the implementation of high-fidelity entangling gates between molecules \cite{Ni2018,Hughes2020}.

\begin{acknowledgments}
We thank R. V. Brooks, I. Forbes, and K. K. Roice for early experimental assistance, A. L. Tao and G. Murray for assistance with the frequency stabilisation of the spectroscopy laser,  and H. J. Williams and O.~Dulieu for helpful discussions. We acknowledge support from the UK Engineering and Physical Sciences Research Council (EPSRC) Grants EP/P01058X/1 and EP/V047302/1, UK Research and Innovation (UKRI) Frontier Research Grant EP/X023354/1, the Royal Society and Durham University. The data presented in this paper are available from \url{http://doi.org/10.15128/r19w032304n}.

\end{acknowledgments}
\
\appendix

% \bibliography{Mergo_ref}
%

\end{document}

\title{Supplemental Material:\\
Formation of ultracold molecules by merging optical tweezers}

% author issues fixed by specifiying superscriptaddress in documentclass
\newcommand{\physics}{Department of Physics and Joint Quantum Centre (JQC) Durham-Newcastle, Durham University, South Road, Durham, DH1 3LE, United Kingdom}
\newcommand{\chemistry}{Department of Chemistry and Joint Quantum Centre (JQC) Durham-Newcastle, Durham University, South Road, Durham, DH1 3LE, United Kingdom}

\author{Daniel K. Ruttley}
\thanks{D. K. R. and A. G. contributed equally to this work.}
\affiliation{\physics}
\author{Alexander Guttridge}
\thanks{D. K. R. and A. G. contributed equally to this work.}
\affiliation{\physics}
\author{Stefan Spence}
\affiliation{\physics}
\author{Robert~C.~Bird}
\affiliation{\chemistry}
\author{C. Ruth Le Sueur}
\affiliation{\chemistry}
\author{Jeremy M. Hutson}
\email{j.m.hutson@durham.ac.uk}
\affiliation{\chemistry}
\author{Simon L. Cornish}
\email{s.l.cornish@durham.ac.uk}
\affiliation{\physics}

\maketitle

\section*{Coupled-channel bound-state calculations}
\subsection{Near-threshold bound states}

To calculate the energies of near-threshold states of RbCs in a magnetic field, in the absence of tweezer confinement, we carry out coupled-channel calculations that take full account of the electron and nuclear spins of the two atoms. The Hamiltonian for the interacting pair is
\begin{equation}
\label{full_H}
\hat{H} =\frac{\hbar^2}{2\mu}\left[-R^{-1}\frac{d^2}{dR^2}R
+\frac{\hat{L}^2}{R^2}\right]+\hat{H}_\textrm{Rb}+\hat{H}_\textrm{Cs}+\hat{V}(R),
\end{equation}
where $R$ is the internuclear distance, $\mu$ is the reduced mass, and $\hbar$ is the reduced Planck constant. $\hat{L}^2$ is the angular momentum operator for relative motion of the atoms. The single-atom Hamiltonians $\hat{H}_A$ contain the hyperfine couplings and the Zeeman interaction with the magnetic field. The interaction operator $\hat{V}(R)$ contains the two isotropic Born-Oppenheimer potentials, for the $X^1\Sigma_g^+$ singlet and $a^3\Sigma_u^+$ triplet states, and anisotropic spin-dependent couplings which arise from dipole-dipole and second-order spin-orbit coupling. In the present work we use the interaction potential of Takekoshi \emph{et al.}\ \cite{Takekoshi2012}, which was fitted to extensive experimental results from Fourier transform (FT) molecular spectroscopy, Feshbach resonances and near-threshold bound states.

The total wavefunction is expanded as
\begin{equation}
\Psi(R,\xi) = R^{-1} \sum_j \psi_j(R) \Phi_j(\xi),
\end{equation}
where $\{\Phi_j(\xi)\}$ represent a basis set that spans all coordinates except $R$, collectively denoted $\xi$. In the present work we use a coupled-atom basis set, with basis functions
\begin{equation}
|(s_\textrm{Rb} i_\textrm{Rb})f_\textrm{Rb} m_\textrm{Rb}\rangle
|(s_\textrm{Cs} i_\textrm{Cs})f_\textrm{Cs} m_\textrm{Cs}\rangle
|L M_L\rangle.
\label{eq:basis-part}
\end{equation}
Here $s_A$ and $i_A$ are the electron and nuclear spins of atom $A$, $f_A$ is its total angular momentum, $m_A$ is the projection of $f_A$ onto the axis of the magnetic field, and $L$ and $M_L$ are the quantum numbers for the molecular rotation and its projection.
Substituting this expansion into the total Schr\"odinger equation produces a set of coupled differential equations that can be solved to obtain either bound-state or scattering properties. The parity $(-1)^L$ and total projection $M_\textrm{tot}=m_\textrm{Rb}+m_\textrm{Cs}+M_L$ are conserved quantities, so separate calculations are carried out for each value of them required. In the present work we use a basis set that contains all allowed spin and rotational functions of the form (\ref{eq:basis-part}) for the required $M_\textrm{tot}$ and even parity, limited here by $L_{\rm max}=4$ to allow the calculation of g-wave states.

We carry out bound-state calculations using the packages \textsc{bound} and \textsc{field} \cite{bound+field:2019,mbf-github:2022}. These packages locate bound states as a function of energy at fixed field, or as a function of field at fixed energy, respectively. They propagate solutions of the coupled-channel equations for a trial energy (or field) from short range and from long range to a matching point $R_\textrm{mid}$ in the classically allowed intermediate region, and then converge upon energies (or fields) at which the wavefunction and its derivative are continuous at $R_\textrm{mid}$. The propagators, step sizes and other numerical methods used are similar to those used in Ref.\ \cite{Brookes:2022}, so will not be described in detail here.

\subsection{Two atoms in overlapping tweezers}

To build a quantitative understanding of the mergoassociation process, we consider two atoms with masses $m_1$ and $m_2$ confined in adjacent tweezers, separated by a vector $\Delta \boldsymbol{z}$. The potentials of the traps are approximated as harmonic with frequencies $\omega_1$ and $\omega_2$. The atoms are treated as unstructured. The motion may be factorized approximately into terms involving the relative and center-of-mass coordinates of the pair, $\boldsymbol{R}$ and $\mathbfcal{R}$ respectively.  The frequencies for relative and center-of-mass motion due to the trap potentials are \cite{Deuretzbacher:2008}
\begin{eqnarray}
\omega_\textrm{rel} &= \sqrt{\left({m_2\omega_1^2+m_1\omega_2^2}\right)\big/\left(m_1+m_2\right)},
\label{eq:wrel}\\
\omega_\textrm{com} &= \sqrt{\left(m_1\omega_1^2+m_2\omega_2^2\right)/\left(m_1+m_2\right)}
\label{eq:conf-freq}
\end{eqnarray}
together with a coupling term between the two that is proportional to $\Delta\omega^2 = \omega_1^2-\omega_2^2$. The relative and center-of-mass motions are thus uncoupled if the atoms are identically trapped. If the confinement is harmonic but non-spherical, these equations apply separately along each principal axis $x$, $y$, $z$ of the trap.

In the present work we consider a model system in which the two tweezers are individually spherically symmetric, and each atom is unaffected by the laser that traps the other atom. We neglect coupling between the relative and center-of-mass motions. This is the same model as treated by Stock \emph{et al.}\ \cite{Stock2003}, but we generalize it to nonidentical atoms and use a different method of solution.
The potential for relative motion due to the combined traps, $V_\textrm{rel}^\textrm{trap}(\boldsymbol{R})$, is centered at
$\boldsymbol{R}=\Delta\boldsymbol{z}$,
\begin{equation}
V_\textrm{rel}^\textrm{trap}(\boldsymbol{R}) = \frac{1}{2}\mu\omega_\textrm{rel}^2 |(\boldsymbol{R}-\Delta\boldsymbol{z})|^2.
\label{eq:Vspher}
\end{equation}

We choose the $z$ axis to lie along $\Delta\boldsymbol{z}$ and expand the potential as
\begin{equation}
V_\textrm{rel}^\textrm{trap}(\boldsymbol{R}) = \sum_{\lambda} V_{\lambda}(R) P_{\lambda}(\cos\theta),
\end{equation}
where $R,\theta,\phi$ are the polar coordinates of $\boldsymbol{R}$ and $P_{\lambda}(\cos\theta)$ is a Legendre polynomial. For the potential (\ref{eq:Vspher}), only terms with $\lambda=0$ and 1 exist. These are
\begin{align}
V_{0}(R)&=\frac{1}{2}\mu\omega_\textrm{rel}^2 [(\Delta z)^2 + R^2]; \\
V_{1}(R)&=-\mu\omega_\textrm{rel}^2 R \Delta z,
\end{align}
where the constant term involving $(\Delta z)^2$ is included to place the trap minimum at zero energy.
The potential may also be expressed in terms of the harmonic length for relative motion, $\beta_\mathrm{rel}=[\hbar/(\mu\omega_\textrm{rel})]^{1/2}$.

The Schr\"odinger equation for relative motion is
\begin{align}
\Biggl[\frac{\hbar^2}{2\mu} &\left(-R^{-1}\frac{d^2}{dR^2}R + \frac{\hat{L}^2}{R^2} \right)\nonumber\\ &+V_\textrm{rel}^\textrm{trap}(\boldsymbol{R}) + V_\textrm{int}(\boldsymbol{R}) - E \Biggr]\Psi(R,\theta,\phi) =  0,
\label{eq:Schrod0}
\end{align}
where $\hat{L}^2$ is the angular momentum operator for relative motion of the atoms and $E$ is
the total energy. The total potential energy is $V(\boldsymbol{R}) = V_\textrm{rel}^\textrm{trap}(\boldsymbol{R}) + V_\textrm{int}(\boldsymbol{R})$, where $V_\textrm{int}(\boldsymbol{R})$ is the interaction potential between the two atoms.
To solve Eq.\ \ref{eq:Schrod0}, we expand the wavefunction as
\begin{equation}
\Psi(R,\theta,\phi) = R^{-1} \sum_{L} \psi_{LM}(R) Y_{LM}(\theta,\phi),
\label{eq:psiexp0}
\end{equation}
where $Y_{LM}(\theta,\phi)$ are spherical harmonics normalized to unity, and the projection quantum number $M$ is a conserved quantity. Substituting the expansion (\ref{eq:psiexp0}) into Eq.\ \ref{eq:Schrod0} gives a set of coupled equations for the
channel functions $\psi_{LM}(R)$,
\begin{equation}\frac{d^2\psi_{LM}}{dR^2}
=\sum_{L'}\left[W_{LL'M}(R)-{\cal E}\delta_{LL'}\right]\psi_{L'M}(R),
\end{equation}
where $\delta_{ij}$ is the Kronecker delta, ${\cal E}=2\mu E/\hbar^2$ and the contribution of $V_\textrm{rel}^\textrm{trap}(\boldsymbol{R})$ is
\begin{align}
W_{LL'M}(R) = &\frac{L(L+1)}{R^2} \delta_{LL'}+ \frac{2\mu}{\hbar^2}V_0(R) \delta_{LL'} \nonumber\\
&+ \frac{2\mu}{\hbar^2}V_1(R) (-1)^M [(2L+1)(2L'+1)]^{1/2}
\nonumber\\ &\qquad\times
\left(\begin{matrix} L & 1 & L' \cr -M & 0 & M \end{matrix} \right)
\left(\begin{matrix} L & 1 & L' \cr  0 & 0 & 0  \end{matrix} \right).
\end{align}

We once again solve these coupled equations using \textsc{bound} \cite{bound+field:2019,mbf-github:2022}. Parity is not conserved, so functions for both even and odd $L$ are included. The size of the basis set required increases with $\Delta z$, but including functions up to $L_\textrm{max}=18$ gives convergence of the energies to 6 significant figures for the largest $\Delta z$ considered here.
 
Since $\Delta z$ is much larger than the range of the interaction potential $V_\textrm{int}(\boldsymbol{R})$, we represent $V_\textrm{int}(\boldsymbol{R})$ as a point contact potential that reproduces the s-wave scattering length $a_\textrm{s}$.
The effect of this is to impose a boundary condition on the log-derivative of the wavefunction, \begin{equation}
\frac{d\psi_{00}}{dR} [\psi_{00}(R)]^{-1}=-1/a_\textrm{s}
\end{equation}
at $R=0$. A contact potential has no effect on the bound states or scattering for $M\ne0$, so calculations are performed only for $M=0$. The coupled equations are propagated from 0 to $R_\textrm{match}\approx\Delta z$ using the fixed-step symplectic log-derivative propagator of Manolopoulos and Gray \cite{MG:symplectic:1995} with a step size of 25~\AA\ and from $R_\textrm{max}\approx\Delta z + 5\beta_\textrm{rel}$ to $R_\textrm{match}$ using the variable-step Airy propagator of Alexander and Manolopoulos \cite{Alexander:1987}.

This approach differs from that of Stock \emph{et al.}\ \cite{Stock2003} in that it does not need basis sets for the internuclear distance $R$, which is handled efficiently by the propagation. The $R$-dependent coupling matrices in our formulation have dimension $L_\textrm{max} + 1$, so are much smaller than the Hamiltonian matrix in a basis set that includes radial functions.

\begin{figure}[tbh]
\includegraphics[width=\columnwidth]{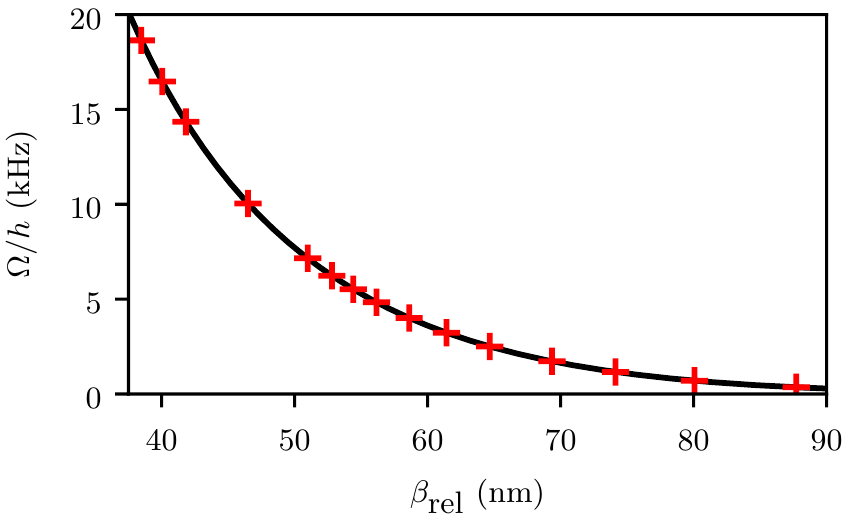}
\caption{Effective coupling matrix element $\Omega$ as a function of confinement length $\beta_\textrm{rel}$ for relative motion, calculated for RbCs with a point-contact interaction potential corresponding to $a_\textrm{s}=645\ a_0$. Red crosses indicate calculated points.}
\label{fig:crossing}
\end{figure}

The coupled-channel calculations allow us to produce energy-level diagrams such as that shown in Fig.\ 1(a) of the main text for any values of $a_\textrm{s}$, $\Delta z$ and $\beta_\textrm{rel}$. In each case there is an avoided crossing between the molecular bound state and the lowest confined atom-pair state at $\Delta z_{_{\textrm{X}}}$, which is approximately $\beta_\textrm{rel} \sqrt{3 + \beta_\textrm{rel}^{2} / a_\textrm{s}^{2}}$. We locate this crossing and then determine its precise position and effective coupling matrix element $\Omega$ by a local fit of the state energies near $\Delta z=\Delta z_{_{\textrm{X}}}$ to the eigenvalues of a $2\times2$ matrix
\begin{equation}
\left(
\begin{matrix}
E_\textrm{X} + d_\textrm{mol}(\Delta z-\Delta z_{_{\textrm{X}}}) & \Omega \\
\Omega & E_\textrm{X} + d_\textrm{at}(\Delta z-\Delta z_{_{\textrm{X}}})
\end{matrix}
\right),
\end{equation}
where $E_\textrm{X}$ is the central energy of the avoided crossing.
Fig.~\ref{fig:crossing} shows the resulting values of $\Omega$ as a function of $\beta_\textrm{rel}$ for RbCs with $a_\textrm{s}=645\ a_0$. It is an expanded and more detailed version of Fig.\ 1(c) of the main text. To a good approximation, the gradients of the atom-pair and molecular states with respect to $\Delta z$ at $\Delta z_{_{\textrm{X}}}$ are $d_\textrm{at}=0$ and
\begin{equation}
d_\textrm{mol}=\mu\omega_\textrm{rel}^2 \Delta z_{_{\textrm{X}}} = \frac{\hbar^2 \Delta z_{_{\textrm{X}}}}{\mu\beta_\textrm{rel}^4}.
\end{equation}

The semiclassical probability of traversing the avoided crossing adiabatically is obtained from the Landau-Zener formula,
\begin{equation} \label{eq:LZ}
    P_\textrm{LZ} = \exp{\left(\frac{-2\pi\Omega^2}{\hbar \left|(d_\textrm{mol}-d_\textrm{at})\frac{d\Delta z}{dt}\right|}\right)}.
\end{equation}

\section*{Preparation of R\lowercase{b}+C\lowercase{s} atom pairs in the motional ground state}
Each experimental run starts by loading two 1D arrays of optical tweezers which trap Rb and Cs atoms. 
Rearrangement is performed on each of these 1D arrays in order to prepare a single Rb and single Cs atom in spatially separated optical tweezers at wavelengths $817$~nm and $1065$~nm respectively. 
At the foci, the beam waists are $\left\{w^{817}_1, w^{817}_2\right\} = \left\{0.82(1),0.92(1)\right\}$~\textmu m and $\left\{w^{1065}_1, w^{1065}_2\right\} = \left\{1.05(1),1.16(1)\right\}$~\textmu m for the merging and non-merging axes respectively.

The initial preparation of Rb and Cs in species-specific tweezers has been described in detail in Ref.~\cite{brooks21}. We prepare and initially image a single Rb atom alongside a single Cs atom in $68.6(4)$\% of experimental runs.
Only these runs are post-selected for further analysis.
The Rb and Cs atoms are then cooled to the motional ground state of their traps with Raman sideband cooling before being prepared in the hyperfine states $(f=1,m_f=1)$ and $(3,3)$ respectively, as described in Ref.~\cite{Spence22}.

\section*{R\lowercase{b}C\lowercase{s} Optical Spectroscopy}

\begin{figure}[t]
\includegraphics[width=\columnwidth]{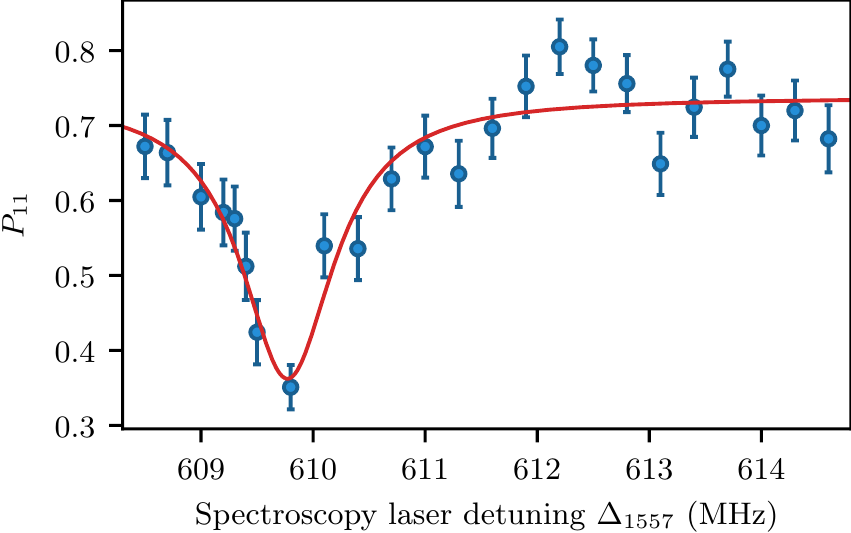}
\caption{Molecule loss caused by driving the transition $a^3\Sigma^+ \rightarrow {^3\Pi_1}$.
When the laser frequency is resonant with the transition, molecules are lost from the $a^3\Sigma^+$ manifold and there is a decrease in the probability $P_{11}$ of reimaging the atom pair after the reversing the association stages of the experimental routine.
\label{fig:1557_lineshape}}
\end{figure}

The RbCs bound-state spectrum presented in the main text is measured with one-photon spectroscopy with light at $1557$~nm from a diode laser (the ``spectroscopy laser'', Toptica, DL pro).
The frequency of this light for the mergoassociation spectroscopy presented in Figure~3 of the main text is $192572.11(4)$~GHz, as measured by a Bristol Instruments 671A-NIR wavemeter with a quoted accuracy of $\pm0.2$ parts per million.
This light is linearly polarised and  drives $\pi$ transitions from the weakly bound molecular states with high $a^3\Sigma^+$ character that we can access with a combination of mergoassociation and magnetic field ramps to the excited state \intermediate from which the molecule can subsequently decay to a number of different states.
At the end of an experimental run we reverse the magnetic field ramps and subsequently unmerge the traps which dissociates any molecules in the state $\mathrm{S} \equiv \,$\one into an atom pair that can be reimaged.
Here the weakly bound molecular state is labelled as $\left(n\left(f_\textrm{Rb},f_\textrm{Cs}\right)L\left(m_\textrm{Rb},m_\textrm{Cs}\right)\right)$, where $n$ is the vibrational quantum number relative to the supporting threshold and $L$ is the rotational quantum number around the centre of mass.
Hence, excitation to the $^{3}\Pi_{1}$ manifold is observed via an increase in the probability of not reimaging the atom pair, as the molecule is lost to a different state from which it cannot be dissociated.
The lineshape of this transition is shown in Fig.~\ref{fig:1557_lineshape}; by measuring the frequency of the transition, we measure the energy of the RbCs molecule.
The detuning $\Delta_\mathrm{1557}$ of the excitation light is relative to the arbitrary frequency reference used to stabilise the laser frequency; details of the laser stabilisation are described in the following section.

\begin{figure*} 
\includegraphics[width=\hsize]{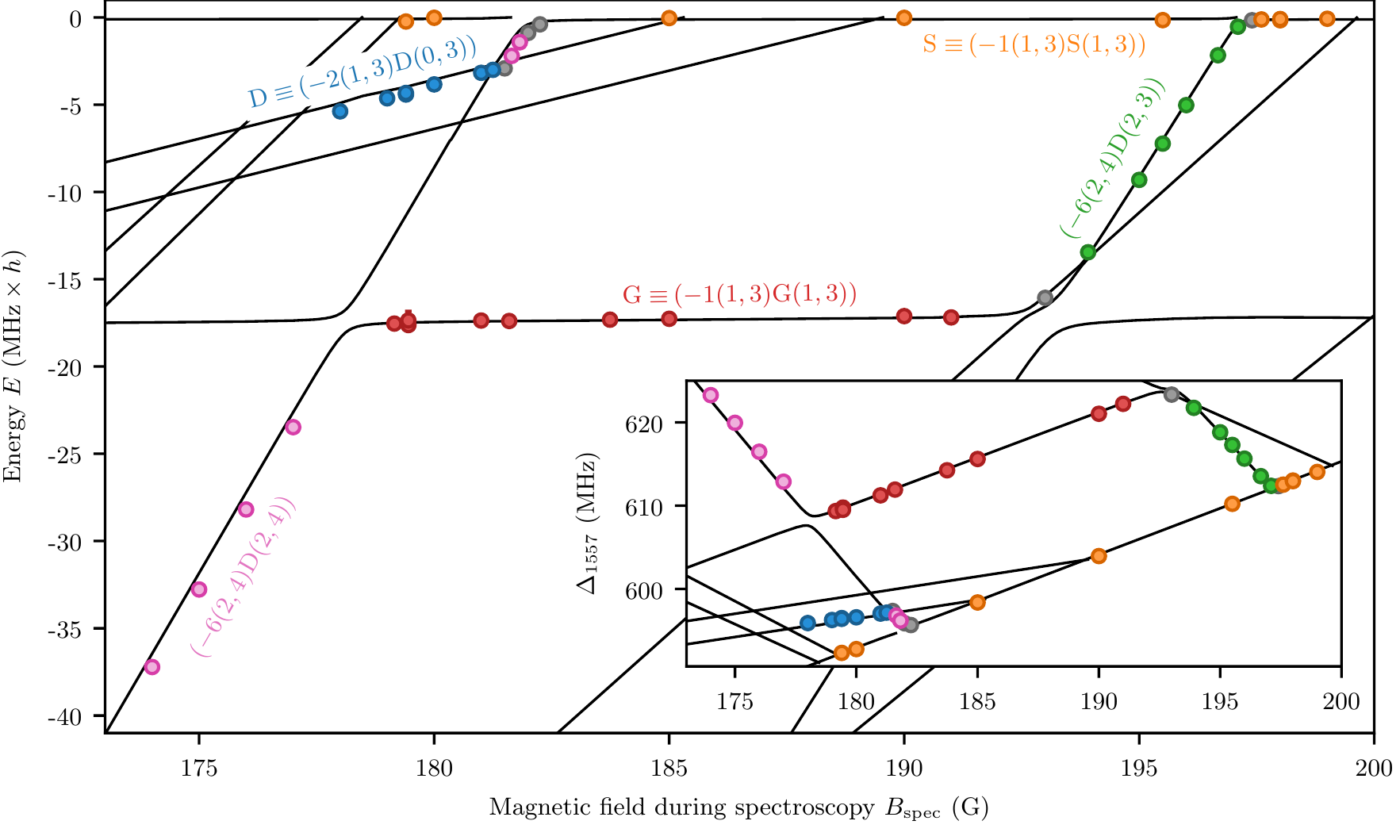}
\caption{Bound-state spectrum (relative to threshold) measured with one-photon spectroscopy after mergoassociation. Colors and their corresponding labels indicate different molecular states labeled as $\left(n\left(f_\textrm{Rb},f_\textrm{Cs}\right)L\left(m_\textrm{Rb},m_\textrm{Cs}\right)\right)$; points in grey indicate measurements in the region of avoided crossings between states. We mergoassociate into the near-threshold state S at various magnetic fields and subsequently navigate the bound-state spectrum with magnetic field sweeps. Black lines are state energies from coupled-channel calculations using the RbCs interaction potential of Ref.~\cite{Takekoshi2012}. The inset shows the same data with energies expressed as the laser detuning $\Delta_\textrm{1557}$ above the arbitrary frequency reference used to stabilise the spectroscopy laser. 
\label{fig:1557_states}}
\end{figure*}

Figure~\ref{fig:1557_states} shows the near-threshold levels of RbCs in the field range of interest here. It is a more detailed version of Fig.\ 2(b) of the main text, with points grouped by the molecular states accessed. The states $\mathrm{D} \equiv \,$\two and $\mathrm{G} \equiv \,$\g in the main text are shown in blue and red respectively.
These two states were chosen for the comparison between mergoassociation and magnetoassociation in the main text because they posses similar transition strengths. 
This is in contrast to the other states with $n=-6$ that have much stronger coupling to the excited state \intermediate which is exploited during two-photon transfer to the rovibrational ground state \cite{Molony2014,Takekoshi2014,Molony2016a}.
The inset shows the detuning of the excitation light from the arbitrary frequency reference.
These data are mapped to molecular energies by fitting the orange points from the state S to the corresponding theoretical energies.
From this we extract the magnetic moment of the excited molecular state $\mu = -0.452(2) \mu_\textrm{B}$, where $\mu_\textrm{B}$ is the Bohr magneton.

\section*{Spectroscopy Laser Stabilisation} \label{sec:laser_setup}
The spectroscopy laser is frequency stabilised with a Pound-Drever-Hall (PDH) technique \cite{Black2001} to an ultra-low expansion cavity (Stable Laser Systems, custom) with finesse $\mathcal{F}=17000(200)$ and free-spectral-range $\omega_\textrm{fsr}=1498.796(3)$~MHz at 1557~nm.
Two sets of frequency sidebands are added to the light going to the cavity by an electro-optic modulator (Thorlabs, LN65S-FC).
One set of sidebands at $\pm19.27$~MHz is used to derive the PDH error signal and the other set is used to tune the frequency offset of the error signal between the cavity transmission peaks to allow for stabilisation to an arbitrary frequency.
Two independent function generators produce the driving tones for these sidebands (Rigol, DG822 and Windfreak Technologies, Synth HD respectively) before they are combined with a power splitter (Minicircuits, ZAPD-2-252-S+) and sent to the modulator.

\begin{figure*}[th]
\includegraphics[width=\hsize]{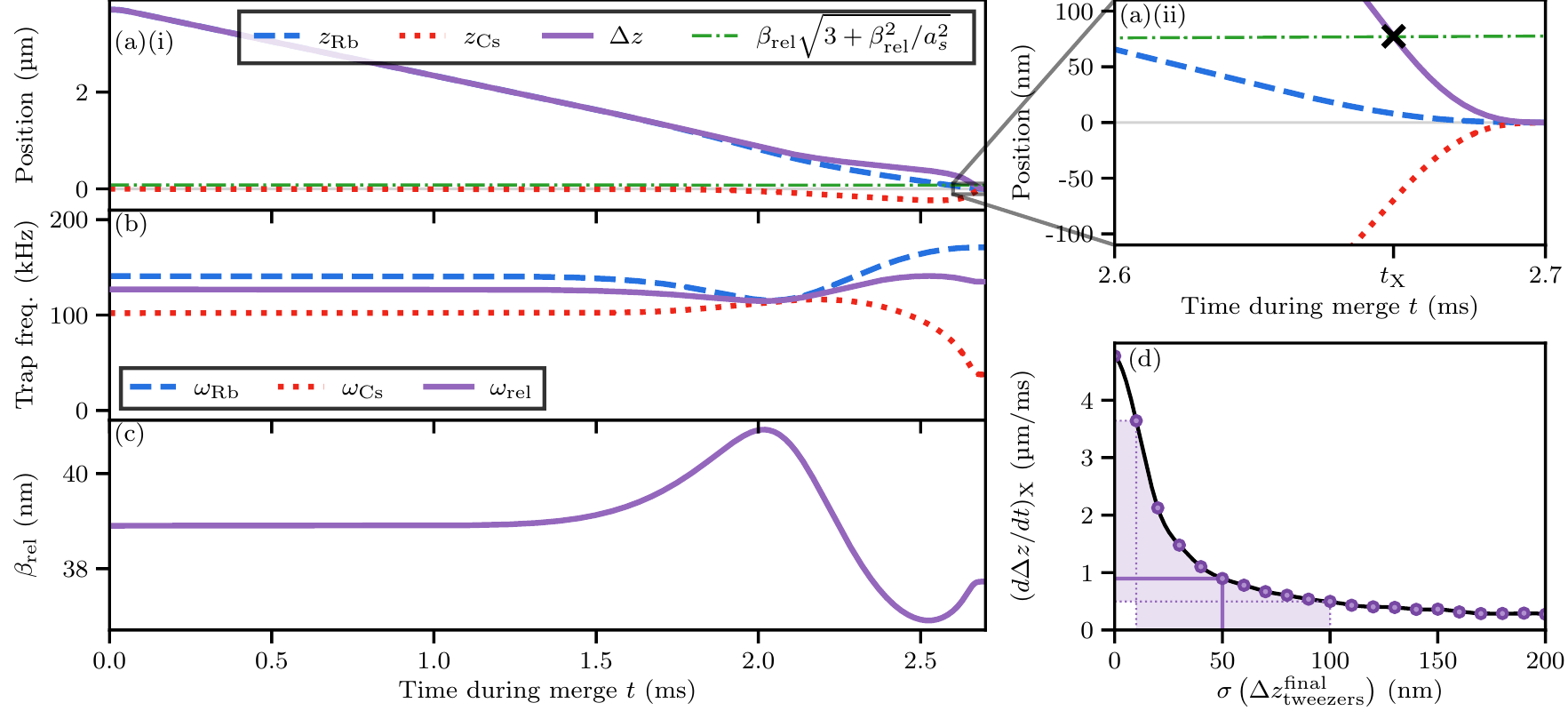}
\caption{
(a)-(c) Simulation results of the merging routine for trap depths corresponding to $\beta_\mathrm{rel}^\mathrm{X} =$ 37.7~nm, finishing with perfectly overlapped tweezers.
(a) The position of the potential minima along the merging axis for the Rb atom, $z_\mathrm{Rb}$, and the Cs atom, $z_\mathrm{Cs}$, and $\Delta z$.
(b) The atomic trap frequencies and the frequency of relative motion.
(c) The harmonic length for relative motion, $\beta_\mathrm{rel}$.
(d) The speed $(d\Delta z/dt)_\mathrm{X}$ at which the avoided crossing is traversed as a function of the standard deviation of tweezer misalignments $\sigma\left(\Delta z_\mathrm{tweezers}^\mathrm{final}\right)$ at the end of the merging routine. 
\label{fig:merging_simulation}}
\end{figure*} 

\section*{Mergoassociation sweeps}

Mergoassociation is performed by sweeping the position of the Rb-trapping tweezer (the ``817~nm tweezer'') to overlap with the Cs-trapping tweezer (the ``1065~nm tweezer'') after the atoms have been cooled to their motional ground state and prepared in the correct hyperfine state.
The position of the 817~nm tweezer is dynamically controlled by deflecting the tweezer beam with a 2D acousto-optic deflector (AA Opto Electronic, DTSXY-400-810) driven with an arbitrary waveform generator (Spectrum Instrumentation, M4i.6631-x8) prior to the focusing objective lens.
The position of the 1065~nm tweezer is controlled with a spatial light modulator (Boulder Nonlinear Systems, PDM512-1064-DVI) before the objective lens; the slow refresh rate ($60$~Hz) precludes the movement of this tweezer during an experimental run.

During the loading, cooling, and state preparation phases, the tweezers are radially separated by approximately $3.7$~\textmu m. 
The $817$~nm tweezer is then swept with a hybrid-jerk trajectory \cite{Liu2019} with hybridicity parameter $\alpha=0.95$ in $2.7$~ms to overlap with the $1065$~nm tweezer.

The combined potential experienced by the atoms from both tweezers during this merging routine is simulated.
We calculate the trajectory of the atoms and the trap frequencies along the merging axis experienced by Rb, $\omega_\textrm{Rb}$, and Cs, $\omega_\textrm{Cs}$.
From these, we obtain the trap frequency for relative motion $\omega_\textrm{rel}$ (\ref{eq:wrel}) and the harmonic length for relative motion $\beta_\mathrm{rel}$ as functions of time. 
We take the avoided crossing to occur when $\Delta z = \beta_\mathrm{rel} \sqrt{3+\beta_\mathrm{rel}^2/a_s^2}$; this occurs at time $t_\mathrm{X}$ with separation $\Delta z_{_{\textrm{X}}}$ and confinement length $\beta_\textrm{rel}^\textrm{X}$.
The results of this simulation for trap depths corresponding to $\beta_\mathrm{rel}^\mathrm{X}=37.7$~nm are shown in Figure \ref{fig:merging_simulation}.

The critical separation at which we traverse the avoided crossing is typically around $\Delta z\sim100$~nm.
This is near the end of the merging routine, where small deviations in the trajectory significantly affect the  speed $(d\Delta z/dt)_\mathrm{X}$ at which the avoided crossing is traversed.
To account for this, we perform Monte Carlo simulations where the final tweezer separation is sampled from a Gaussian distribution with standard deviation $\sigma\left(\Delta z_\mathrm{tweezers}^\mathrm{final}\right)$.
Figure~\ref{fig:merging_simulation}(d) shows the mean speed $(d\Delta z/dt)_\mathrm{X}$ as a function of this standard deviation.
From independent measurements of the drift in tweezer positions over time, we expect the standard deviation to be approximately $50^{+50}_{-40}$~nm; this results in the range of speeds $(d\Delta z/dt)_\mathrm{X} = 0.9^{+2.7}_{-0.4}$~\textmu m/ms used in Fig.~3 of the main text. 
We note that this analysis predicts a most likely speed similar to a much simpler analysis assuming a constant speed throughout the merging process $3.7$~\textmu m$/ 2.7$~ms $\sim 1.4$~\textmu m/ms.

\begin{figure}[t] 
\includegraphics[width=\columnwidth]{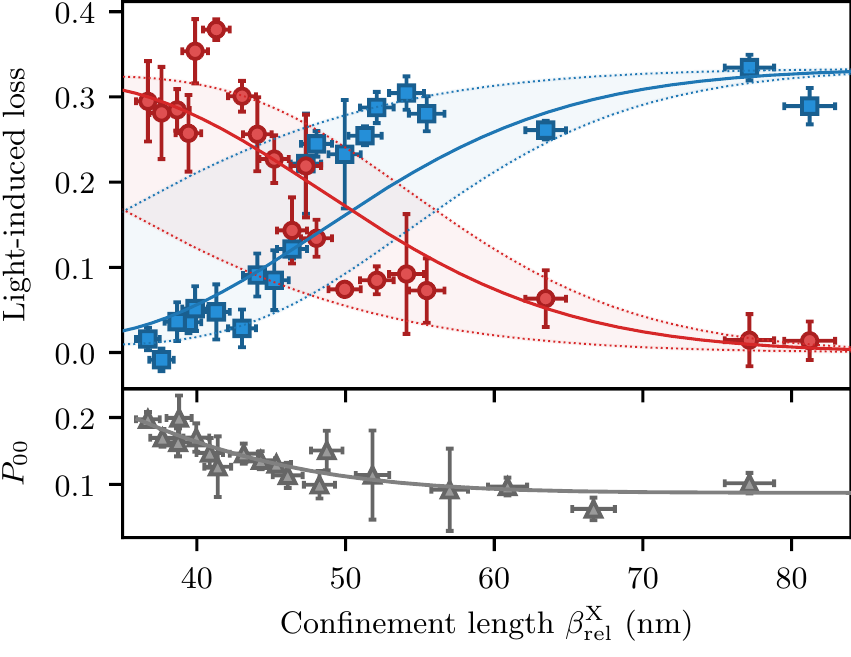}
\caption{Upper panel: The probability of light-induced loss of molecules formed by mergoassociation (red circles) and magnetoassociation (blue squares) as a function of $\beta_\textrm{rel}^\textrm{X}$. 
The theory curves show the calculated Landau-Zener probabilities scaled to match the light-induced loss. 
The shaded regions indicate the experimental uncertainty in the merging speed.
Lower panel: The probability $P_{00}$ of molecule loss in the absence of spectroscopy light. The solid line is a fit to an error function.
\label{fig:unnormalised_loss}}
\end{figure}

Experimentally, when we deliberately misalign the tweezers we observe that the efficiency of mergoassociation drops to half its maximum value for a given confinement strength for tweezer separations $\Delta z_\mathrm{tweezers}^\mathrm{final}\sim 200$~nm.

\section*{Loss of mergoassociated molecules in the absence of spectroscopy light}
Figure~3(b) of the main text displays the amount of loss induced by light with wavelength 1557~nm, tuned to resonance with either the state D or G. 
In Figure~\ref{fig:unnormalised_loss} we present the same experimental data presented in the main text, alongside data with no resonant light pulse applied. 
The grey points in the lower panel show that there is a small probability of detecting zero atoms (molecule loss) when no resonant light is applied, which provides some evidence that the mergoassociation may have a higher probability of molecule formation in our experimental setting than magnetoassociation. 
It is evident that there is a small increase  in the loss probability for stronger confinement, with a similar trend to that observed in the measurement of light-induced loss.
This suggests that mergoassociated molecules are produced that are not excited by the spectroscopy light. 
This is probably caused by loss of the molecules before excitation due to photon scattering. In our experimental routine, molecules formed by mergoassociation spend a greater amount of time in states which are more deeply bound. These states scatter more photons from the trapping light, causing faster loss compared to weakly bound states such as the state S. 

\begin{figure}[t]
\includegraphics[width=\columnwidth]{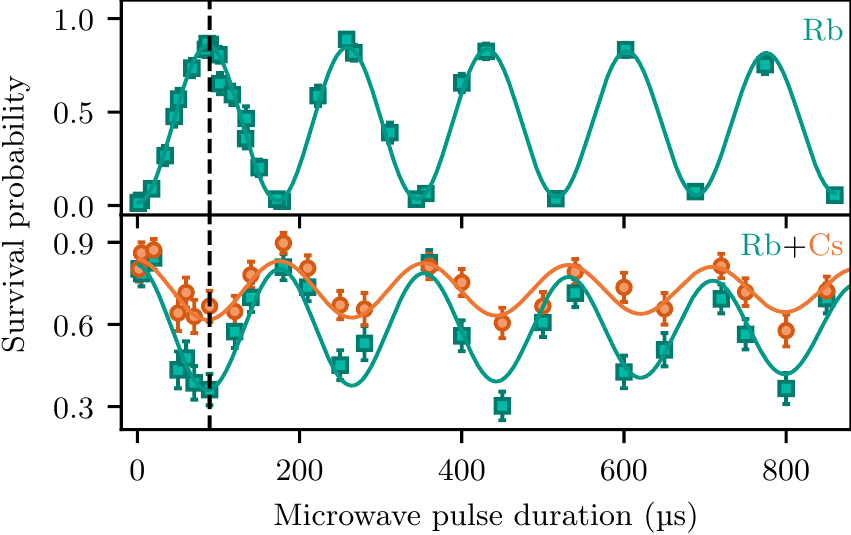}
\caption{Measurement of the Rabi frequencies of the Rb transition $(2,2)\rightarrow(1,1)$ and the RbCs transition $\mathrm{S}\rightarrow\mathrm{S}'$ at 4.78~G. 
Rb atoms in $(f=2)$ are removed after the microwave pulse. 
Green squares (orange circles) show the survival probability of Rb (Cs) atoms.
Upper panel: Rb atoms are prepared in $(2,2)$ before the microwave pulse.
Lower panel: Rb+Cs atom pairs are prepared in $(1,1)+(3,3)$ and mergoassociated into the molecular state $\mathrm{S}$. The microwave pulse is applied at detuning $35$~kHz relative to the Rb transition. Remaining molecules in $\mathrm{S}$ are disassociated back into atom pairs before the removal of Rb $(f=2)$.
The measured Rabi frequencies are $5.81(4)$~kHz and $5.62(3)$~kHz respectively.
For the experiment presented in Fig.~4 of the main text an $89$~\textmu s pulse (black dashed line) is used. 
\label{fig:MW_pulse_times}}
\end{figure}

The observed increase in the molecule formation probability is most likely from a greater probability of preparing the atom pair in the motional ground state when using more confining traps during the merging process. 
This behaviour is explained by heating during the merging process that depends on the intensity of the 817~nm tweezer. The formation of an acoustic cavity in the AOD crystal \cite{Woody2018,Liu2019} leads to parametric modulation of the 817~nm tweezer intensity during merging. A change in the trap confinement during merging therefore leads to a change in the heating experienced by the Rb atom. The change in the heating experienced by Rb is fairly small for the parameters explored here, as evidenced by the observation of magnetoassociation in the absence of mergoassociation (and vice versa). 
Any significant heating would cause a drastic reduction in the population of the motional ground state and therefore the efficiency of molecule formation.

\section*{Microwave spectroscopy}
 We perform microwave spectroscopy to measure molecule formation at a magnetic field of 4.78~G. For experiments in which only a Rb atom is initially prepared, the frequency of the microwave photon is tuned into resonance with the Rb atomic transition $(1,1) \rightarrow (2,2)$. Following the microwave pulse, we apply a pulse of light resonant with Rb atoms in the $(f=2)$ state. This allows spin-sensitive detection of Rb by ejecting  $(f=2)$ atoms from the tweezer. By varying the duration of the resonant microwave pulse we observe the Rabi oscillations shown in the upper panel of  Fig.~\ref{fig:MW_pulse_times}.

 In the lower panel of Fig.~\ref{fig:MW_pulse_times} we show the recapture probabilities of Rb atoms (green squares) and Cs atoms (orange circles) under experimental conditions where we expect mergoassociation of RbCs molecules. The frequency of the microwave photon is detuned by $35(2)$~kHz relative to the Rb transition $(1,1) \rightarrow (2,2)$ in order to drive the molecular transition between states $\mathrm{S}$ and $\mathrm{S}' \equiv \,$\onespinflip. When the microwave frequency is on resonance with the molecular transition, we observe a correlated reduction in the Cs and Rb survival probabilities. In this measurement state-selective detection is applied only to the Rb atom, so the observed change in the Cs survival probability reflects the loss of a Cs atom (whereas the change in the Rb survival probability reflects a change of hyperfine state and/or loss of the atom).  
We do not observe loss of Cs atoms when the microwave frequency is tuned onto resonance with the atomic transition.
This suggests either significant decay of the state $\mathrm{S}'$ for our experimental conditions or failure to break apart the spin-flipped molecule when unmerging the optical tweezers. We note that the binding energy of $\mathrm{S}'$  ($80~\mathrm{ kHz}\times h$) is smaller than that of $\mathrm{S}$  ($122~\mathrm{ kHz}\times h$), so that the latter explanation is less likely.
 
We measure very similar Rabi frequencies for the two transitions: $5.81(4)$~kHz for the atomic transition and $5.62(3)$~kHz for the molecular transition. 
Our chosen pulse time of $89$~\textmu s approximates a $\pi$-pulse for both atomic and molecular transitions. The molecule formation efficiency may then be inferred from the relative depths of the features in Fig.~4 of the main text.

%

% \bibliography{Mergo_ref}